\documentclass{article}
\usepackage{arxiv}

\usepackage[utf8]{inputenc} 
\usepackage[T1]{fontenc}    
\usepackage{hyperref}       
\usepackage{url}            
\usepackage{booktabs}       
\usepackage{amsmath,amssymb,amsfonts}

\usepackage{graphicx}
\usepackage{amsmath}
\usepackage{amssymb}
\usepackage{multirow}
\usepackage{subcaption}
\usepackage{booktabs}
\usepackage{listings}
\usepackage{xcolor}
\usepackage{multirow}
\usepackage{url}
\usepackage{longtable}
\usepackage{makecell, cellspace, caption}
\usepackage{array}
\newcolumntype{L}[1]{>{\raggedright\let\newline\\\arraybackslash\hspace{0pt}}m{#1}}
\newcolumntype{C}[1]{>{\centering\let\newline\\\arraybackslash\hspace{0pt}}m{#1}}
\newcolumntype{R}[1]{>{\raggedleft\let\newline\\\arraybackslash\hspace{0pt}}m{#1}}

\title{uARMSolver: A framework for Association Rule Mining}

\author{ \href{https://orcid.org/0000-0002-9964-6957}{\includegraphics[scale=0.06]{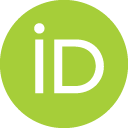}\hspace{1mm}Iztok Fister} \\
	University of Maribor\\
	\texttt{iztok.fister@um.si} \\
	\And
	\href{https://orcid.org/0000-0002-6418-1272}{\includegraphics[scale=0.06]{orcid.png}\hspace{1mm}Iztok Fister Jr.} \\
	University of Maribor\\
	\texttt{iztok.fister1@um.si} \\
}

\begin{document}
\maketitle

\begin{abstract}
The paper presents a novel software framework for Association Rule Mining named uARMSolver. The framework is written fully in C++ and runs on all platforms. It allows users to preprocess their data in a transaction database, to make discretization of data, to search for association rules and to guide a presentation/visualization of the best rules found using external tools. As opposed to the existing software packages or frameworks, this also supports numerical and real-valued types of attributes besides the categorical ones. Mining the association rules is defined as an optimization and solved using the nature-inspired algorithms that can be incorporated easily. Because the algorithms normally discover a huge amount of association rules, the framework enables a modular inclusion of so-called visual guiders for extracting the knowledge hidden in data, and visualize these using external tools.
\end{abstract}


\section{Introduction}
\label{sec:introduction}
Association Rule Mining (ARM) belongs to a set of Machine Learning (ML) methods that have been receiving more attention in the ML community. The task of the method is to discover interesting relations between features in transaction databases. These relations are described in the form of implication rules reflecting association between these features. 

The problem was originally defined by Agrawal et al.~\cite{agrawal1993mining}. Since this definition, a lot of algorithms for generating association rules have been proposed in past years. A large portion of these algorithms are based on deterministic methods~\cite{borgelt2005implementation}, while, nowadays, a lot of algorithms are based on stochastic population-based methods~\cite{telikani2020survey}. A very important and positive argument for utilizing these methods lies in their versatility for dealing with both categorical and numerical attributes~\cite{altay2019performance}, which was a huge bottleneck for deterministic methods~\cite{Fister2018ARM}. 

Typically, the algorithms for ARM discover a huge amount of association rules, from which the relevant information is hard to extract and present to users. Although new so-called explaining methods have emerged recently, like visualization~\cite{fister2020information}, working with ARM is anything but easy for the users. In this sense, three main problems have arisen, as follows: data preparation (i.e., preprocessing), implementation of algorithms for ARM, and, finally, interpretation of the results. Usually, people confronted with solving this problem in practice do not have enough experience and also knowledge about how the algorithms work. 

Therefore, the purpose of this study is to develop a universal framework for ARM, which includes all three steps for solving this problem, i.e., preprocessing, optimization, and visualization. The main advantages of the framework are: speed, modular design, and open-source coding. The speed is ensured by developing this in a programming language C++. The modular design means that the new components, like algorithms for ARM or visualization methods, can easily be added to the framework. Open-source refers to code that people can modify and share publicly due to placing this on GitHub~\cite{GitSite}.

The structure of the remainder of the paper is as follows: Section~\ref{sec_2} defines the problem of ARM formally. A description of the Universal ARM Solver (uARMSolver) is a subject of Section~\ref{sec_3}. Examples of use are shown in Section~\ref{sec_4}, while the results of experiments are illustrated in Section~\ref{sec_5}. The paper is concluded with Section~\ref{sec_6}, where the performed work is summarized and directions for the future work are outlined.

\section{Problem formulation}~\label{sec_2}
The purpose of this section is to supply to a potential reader with the basic information needed for understanding the subjects that follow. In line with this, the following topics are discussed:
\begin{itemize}
    \item Association Rule Mining,
    \item Numerical Association Rule Mining,
    \item existing software libraries.
\end{itemize}
In the remainder of the paper, all three subjects are illustrated in detail.

\subsection{Association rule mining}
The ARM problem is defined formally as follows: Let us suppose a set of objects $O=\{o_1, \ldots,o_m\}$ and transaction database $D$ are given, where each transaction $T$ is a subset of objects $T \subseteq O$. Thus, the variable $m$ designates the number of objects. Then, an association rule can be defined as an implication:
\begin{equation}
X \Rightarrow Y, 
\label{arule}
\end{equation}
\noindent where $X \subset O$, $Y \subset O$, in $X \cap Y = \emptyset$. The following two measures are defined for evaluating the quality of the association rule~\cite{agrawal1994fast}:
\begin{equation}
\mathit{conf}(X \Rightarrow Y) = \frac{n(X \cup Y)}{n(X)}, 
\label{Eq:1}
\end{equation}
\begin{equation}
\mathit{supp}(X \Rightarrow Y) = \frac{n(X \cup Y)}{N}, 
\label{Eq:2}
\end{equation}
\noindent where $\mathit{conf}(X \Rightarrow Y)\geq C_{\mathit{min}}$ denotes confidence and $\mathit{supp}(X \Rightarrow Y) \geq S_{\mathit{min}}$ support of association rule $X \Rightarrow Y$. There, $N$ in equation~(\ref{Eq:2}) represents the number of transactions in the transaction database $D$, and $n(.)$ is the number of repetitions of the particular rule $X \Rightarrow Y$ within $D$. Additionally, $C_{\mathit{min}}$ denotes minimum confidence and $S_{\mathit{min}}$ minimum support, determining that only those association rules with confidence and support higher than $C_{\mathit{min}}$ and $S_{\mathit{min}}$ are taken into consideration, respectively.

\subsection{Numerical association rule mining}
Numerical Association Rule Mining (NARM) extends the idea of ARM, and is intended for mining association rules where attributes in a transaction database are represented by numerical values. Usually, traditional algorithms, e.g. Apriori, require a discretization of numerical attributes before they are ready to use. The discretization is sometimes trivial, and sometimes does not have a positive influence on the results of mining. On the other hand,  many methods for ARM exist that do not  require the discretization step before applying the process of mining. 

Most of these methods are based on population-based nature-inspired metaheuristics, such as, for example, Differential Evolution or Particle Swarm Optimization.  NARM has recently also been featured in some review papers~\cite{altay2019performance,telikani2020survey} which emphasize its importance in the data revolution era.

Each numerical attribute is determined by an interval of feasible values limited by its lower and upper bounds. The broader the interval, the more association rules mined. The narrower the interval, the more specific relations between attributes are discovered. Introducing intervals of feasible values has almost two effects on the optimization: To change the existing discrete search space to continuous, and to adapt these continuous intervals to suite the problem of interest better.

Mined association rules can be evaluated according to several criteria, like support and confidence. However, these cover only one side of the coin. If we would also like to discover the other side, additional measures must be included into the evaluation function.

\subsection{Existing ML software toolboxes}
There is no doubt that ARM is a very popular method in data science. Therefore, some software tools exist for utilizing this task. Table~\ref{tab:0} presents a short overview of software packages for supporting ARM. 

\begin{table}
    \centering
    \caption{Open source software toolboxes for ARM}
    \label{tab:0}
    \begin{tabular}{|L{3cm}|L{3cm}|C{1.2cm}|}
    \hline
    Software toolbox & Implemented in & Reference \\ \hline
    KEEL             & Java      & \cite{alcala2009keel} \\ \hline
    WEKA            & Java          & \cite{hall2009weka} \\ \hline
    Orange           & Python   & \cite{demsar2013orange} \\ \hline
    ARules & R &  \cite{hahsler2011arules} \\ \hline
    ARules & Julia & \cite{stey2020arules} \\ \hline
    \end{tabular}
\end{table}

Knowledge Extraction based on Evolutionary Learning (KEEL) is a software tool for assessing evolutionary algorithms for ML problems including regression, classification, unsupervised learning, ARM etc. It includes several different paradigms, like Pittsburgh, Michigan and Inverse Reinforcement Learning (IRL). The Weikato Environment for Knowledge Analysis (WEKA) allows researchers easy access to state-of-the-art ML methods using a unified workbench. Orange is a ML and data mining toolbox for data analysis using Python scripting and visual programming. It offers a huge number of extension libraries related to ML, like Scikit-learn~\cite{pedregosa11a2012scikit}. The R package ARules implements the basic infrastructure for creating and manipulating transaction databases. It bases on traditional algorithms for ARM (i.e., Apriori and Eclat). The arules ecosystem enables integration of existing ARM methods, and extends the further analysis of mined association rules using functionalities of R. The similar association rule learning package in Julia was developed by the Center for Biomedical Informatics at Brown University (BCBI).

As can be seen from the mentioned review, the majority of ARM implementations are part of open source software toolboxes (e.g., KEEL, Weka and Orange), while packages also exist fully devoted for ARM tasks (e.g., arules). However, the main weakness of the solutions is that these allow primarily working with the categorical attributes, while handling with numerical attributes is not supported in general in all frameworks. On the other hand, all toolboxes are written in higher programming languages. Although they offer interfaces with lower level languages, like C/C++, the pure C/C++ framework does not exist at the moment.

\section{{uARMSolver} framework}~\label{sec_3}
\begin{figure}[htb]
    \centering
    \includegraphics[width=.48\textwidth]{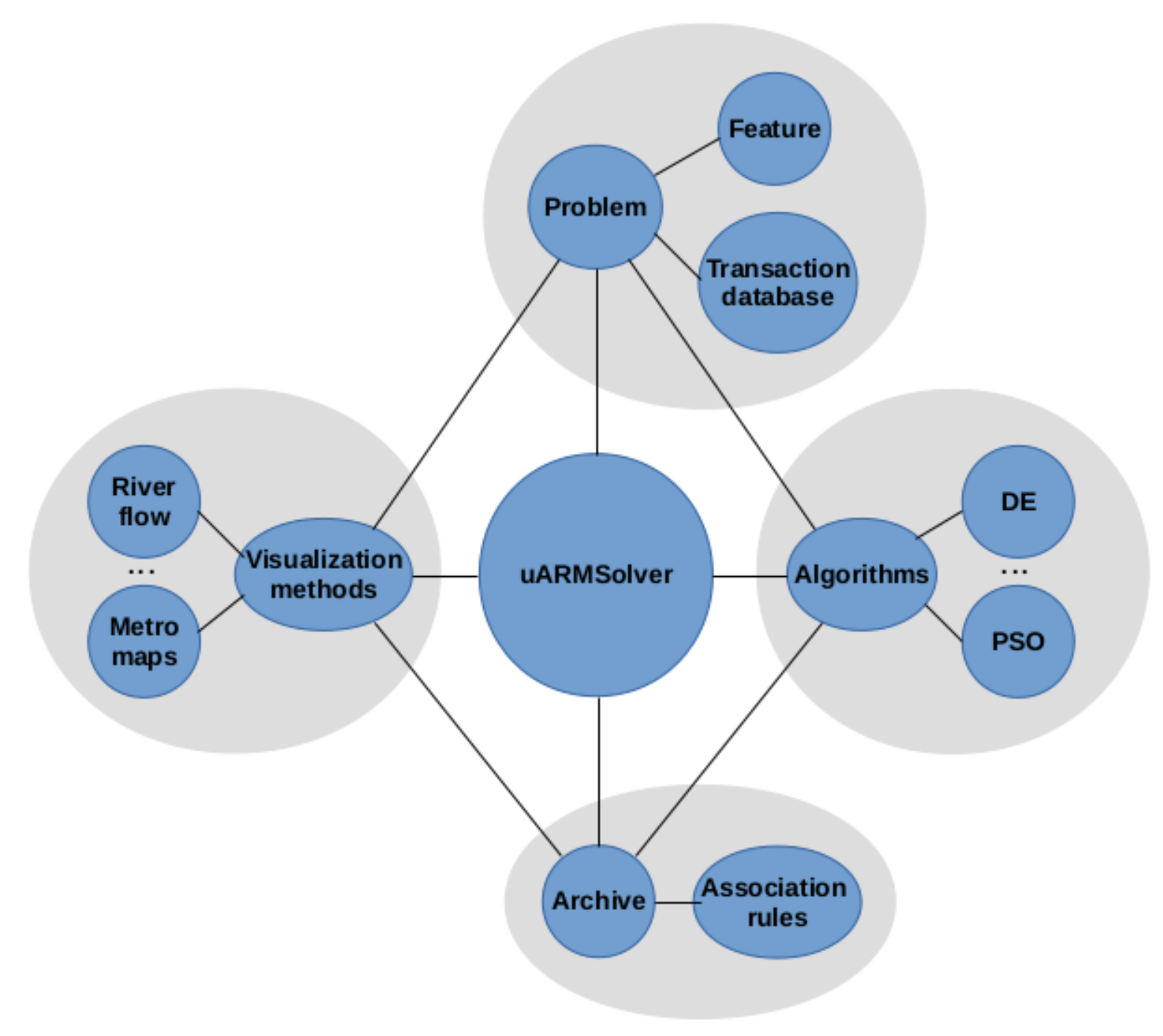}
    \caption{Concept of universal ARM solver.}
    \label{fig:concept}
\end{figure}

The purpose of a uARMSolver is to preprocess input datasets, to make discretization of attributes, to mine association rules using various nature-inspired algorithms, and to guide visualization of the mined association rules. It is designed as a framework, into which various, different components covering each three steps can be integrated. The framework consists of the following four components placed around the core program (Fig.~\ref{fig:concept}):
\begin{itemize}
    \item problem definition,
    \item ARM algorithm selection,
    \item visualization method selection,
    \item association rule archive.
\end{itemize}

The problem definition encompasses the preprocessing step of the ARM, where data saved into datasets are parsed and stored into the transaction database. The supported format of the datasets is similar to those used in UCI Machine Learning Repository datasets~\cite{uci2020repository}, where the first line describes features, while all the following lines define transactions, i.e., sequences of attributes delimited by period character (','). Actually, each $i$-th attribute in the line designates a pair <feature$_i$,attribute$_i$> using the feature laying at the same position. This pair can emerge in the mined association rule in the form of ''feature$_i$\_attribute$_i$''.

The proposed framework is open to including various nature-inspired algorithms, treating the ARM as an optimization problem. At the moment, two algorithms are included, i.e., Differential Evolution (DE), and Particle Swarm Algorithm (PSO). However, the developers of this framework also plan to support the other algorithms from NiaPy library~\cite{niapy2018vrbancic}.

The nature-inspired algorithms generate a lot of association rules that are hard to understand by ordinal users. In order to explain how the AI-powered systems, like uARMSolver, make decisions by solving the ARM, the so-called Explainable Artificial Intelligence (EAI) has emerged~\cite{gunning2017explainable}. This tries to create more explainable AI models and, thus, enables users to understand, trust, and manage the emerging platform of AI effectively. As a result, more post-hoc explainable techniques have been developed that are capable of analyzing the results of ARM solvers' posterior. Indeed, a visual explanation presents one of the more useful methods. 

The proposed framework supports the concepts of EAI, and, in this sense, developers search for the more suitable visual methods for ARM visualization, e.g., river flow~\cite{Fister2020River} and metro maps~\cite{fister2020information}. Indeed, the preliminary work for using these visual methods in practice has already been done, and, thus, developers search for a way to include them into the uARMSolver framework. However, these methods cannot be applied directly, because they need supplemental algorithms to guide the visualization by analyzing the archive of mined association rules.

The association rule archive is a data structure for storing mined association rules. These rules can be generated either by the nature-inspired algorithms for ARM online or traditional ARM algorithms offline. In the latter case, the association rules are read from input files and analyzed by the post-hoc visual explainable methods.  

\subsection{Classes}
\begin{figure}[htb]
    \centering
    \includegraphics[width=.48\textwidth]{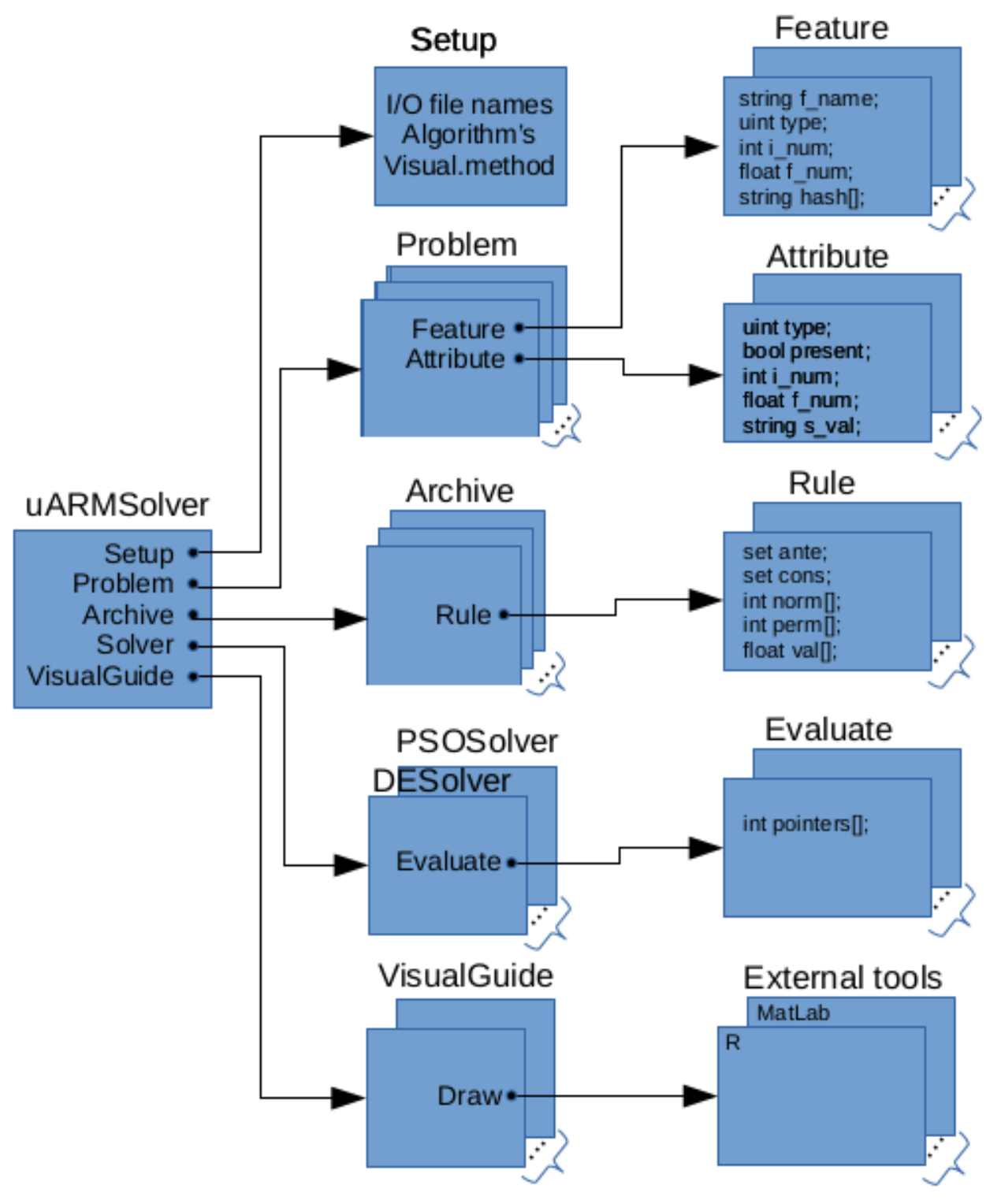}
    \caption{UML diagram of the universal ARM solver.}
    \label{fig:uml}
\end{figure}

C++ is an object-oriented programming language, where each object is defined in classes and instantiated in the main function or within the other classes. The uARMSolver is implemented as a main function instantiating five classes as follows (Fig.~\ref{fig:uml}):
\begin{itemize}
    \item Setup,
    \item Problem,
    \item Archive,
    \item Solver,
    \item VisualGuide.
\end{itemize}
In the remainder of the section, the mentioned classes are illustrated in details.

\subsubsection{Setup class}
The aim of Setup class is to control the behavior of the framework using a setup text file. The setup file is capable of controlling three components (i.e., problem definition, algorithm selection, and visualization method selection). In line with this, it is divided into more independent parameter blocks, where each of these started with a specific reserved word (e.g., \textbf{Problem}, \textbf{DE\_PARAM}, \textbf{FLOW\_PARAM}, etc.), after which definitions of particular parameters follow enclosed between curly brackets. Each parameter is defined within a single line in the following form:
\begin{equation*}
    \text{'parameter\_name'}~=~\text{<token>},
\end{equation*}
where $\text{<token>}$ represents the value (numeric or string) assigned to the parameter.

Setup also enables selection of a specific algorithm, as well as visual method. The former is specified by the \textbf{Algorithm} reserved word, while the latter by the \textbf{Visualization} reserved word. When no algorithm or visualization method are specified, the selection \textbf{NONE} can be used. Contrarily, when the algorithm or visual method are selected, the corresponding parameter block is assigned to them. For instance, the \textbf{DE\_PARAM} block is taken if the DE algorithm is selected. Let us emphasize that there is only one setup file. 

\subsubsection{Problem class}
The Problem class implements data structures and functions for handling with transaction databases. In line with this, it is capable of parsing input data in a UCI ML dataset format. Although the parser was made for parsing the UCI ML datasets, Seaborn data from GitHub~\cite{waskom2020seaborn}, can also be processed with minor changes of the datasets. Obviously, a new parser can easily be included into the framework when needed. The Problem class utilizes two additional classes: Feature and Attribute. The former represents the so-called feature list built after parsing the first line of the dataset, while the latter represents actually the particular transactions consisting of those attributes that belong to a specific one.  Interestingly, the parser distinguishes between three attribute types: categorical (type \textbf{string} in C++), numerical (type \textbf{int} in C++) and real-valued (type \textbf{double} in C++). The type is assigned to the feature according to the context from which it emerged in the transaction database.

In summary, a transaction database is represented as a matrix of dimension $N\times m$, where $N$ is the number of transactions (rows in the matrix) and $m$ the number of attributes in the transaction database (columns in the matrix). 

Indeed, a definition of the Attribute is defined, as illustrated in a following listing.
\begin{lstlisting}
(*@\textbf{public}@*):
    (*@\textbf{uint}@*) (*@\textit{type}@*);    (*@\text{// type}@*)
    (*@\textbf{bool}@*) (*@\textit{present}@*); (*@\text{// present flag}@*)
    (*@\textbf{int}@*) (*@\textit{i\_val}@*);   (*@\text{// integer value}@*)
    (*@\textbf{double}@*) (*@\textit{f\_val}@*);    (*@\text{// real-valued value}@*)
    (*@\textbf{string}@*) (*@\textit{s\_val}@*);    (*@\text{// string value}@*)
\end{lstlisting}
As can be seen from the listing, each attribute is determined by its type, present flag, and value. According to the \textit{type}, the variable of the corresponding C++ type serves for storing value. For instance, the variable \textit{f\_val} is used for storing values of real-valued attributes. The Boolean variable \textit{present} flags, if the attribute is present in the transaction or not.

The Feature class consists of the feature name, type and corresponding domain of values. Let us notice that feature values are actually generalizations of values arising in definite columns. This means that if the type of feature is categorical, the set of discrete attributes is attached to the variable \textit{hash}.  
\begin{lstlisting}
(*@\textbf{public}@*):
    (*@\textbf{string}@*) (*@\textit{f\_name}@*);   (*@\text{// feature name}@*)
    (*@\textbf{uint}@*) (*@\textit{type}@*);    (*@\text{// feature type}@*)
    (*@\textbf{int\_bounds}@*) (*@\textit{i\_num}@*);   (*@\text{// integer bounds}@*)
    (*@\textbf{float\_bounds}@*) (*@\textit{f\_num}@*); (*@\text{// floating-point bounds}@*)
    (*@\textbf{vector<string>}@*) (*@\textit{hash}@*);  (*@\text{// discrete attributes}@*)
\end{lstlisting}
On the other hand, the interval of feasible values are determined, bounded with their minimum and maximum values, when the other types of features are observed.

There can be more Problem objects instantiated from this class, because this framework allows handling of more transaction datasets simultaneously, where each dataset captures data belonging to different periods of time. However, the number of time-dependent transaction datasets are specified using the parameter 'Periods' in the setup file. 

\subsubsection{Archive class}
 Archive class enables working with an archive of association rules. This is defined as a vector of Rule objects, where each object consists of the data structures illustrated in the following listing.
\begin{lstlisting}
(*@\textbf{public}@*):
    (*@\textbf{set <string>}@*) (*@\textit{ante}@*);    (*@\text{// set of antecedents}@*)
    (*@\textbf{set <string>}@*) (*@\textit{cons}@*);    (*@\text{// set of consequent}@*)
    (*@\textbf{vector <int>}@*) (*@\textit{norm}@*);    (*@\text{// normalized vector}@*)
    (*@\textbf{vector <int>}@*) (*@\textit{perm}@*);    (*@\text{// permutation vector}@*)
    (*@\textbf{vector <double>}@*) (*@\textit{val}@*);  (*@\text{// permutation ordering}@*)
\end{lstlisting}
As can be seen from the listing, each association rule from the archive is defined by a set of antecedents and a set of consequents, where elements of both sets are represented in in the form of 'feature\_attribute'. The remaining vectors in the class are used for genotype-phenotype mapping by evaluating the quality of the solution and are discussed in the next subsection.
 
\subsubsection{Solver class}
Solver, illustrated in the following listing, is a more important class and implements an optimization algorithm for ARM. It is designed modularly, and represents a base class from which all the other nature-inspired algorithms are inherited (e.g., DESolver for DE implementation etc.). In line with this, the base class defines two virtual functions, i.e., Setup(.) and Evolve(.). 
\begin{lstlisting}
(*@\textbf{public}@*):
    (*@\text{Solver}@*)((*@\textbf{int}@*) (*@\textit{m}@*), (*@\textbf{int}@*) (*@\textit{Np}@*), (*@\text{Problem}@*) (*@\textit{prob}@*));
    (*@\text{$\sim$Solver}@*)((*@\textbf{void}@*));
    (*@\textbf{virtual void}@*) (*@\text{Setup}@*)((*@\textbf{int}@*) (*@\textit{strat}@*), (*@\textbf{double}@*) (*@\textit{F}@*), (*@\textbf{double}@*)  (*@\textit{CR}@*));
    (*@\textbf{virtual void}@*) (*@\text{Evolve}@*)((*@\textbf{int}@*) (*@\textit{run}@*), (*@\textbf{int}@*) (*@\textit{FEs}@*), (*@\text{Archive}@*) (*@\textit{\&rule}@*));
(*@\textbf{protected}@*):    
    (*@\textbf{vector<double>}@*) (*@\textit{trialSolution}@*); (*@\text{// trial vector}@*)
    (*@\textbf{vector<double>}@*) (*@\textit{popEnergy}@*); (*@\text{// fitness values}@*)
    (*@\textbf{vector<vector <double>>}@*) (*@\textit{pop}@*);  (*@\text{// population}@*)
    (*@\text{Evaluate}@*) (*@\textit{eval}@*);  (*@\text{// fitness function evaluation}@*)
\end{lstlisting}
The first is devoted to setting the parameters of the selected algorithm, while the second launches the evolutionary run using parameters, like the number of the current run, and the maximum number of fitness function evaluations. However, information about the problem to be solved is obtained implicitly by the Solve class during the object creation.

The Solver class also defines data structures for representing a trial solution, fitness values of particular individuals, and the population of individuals. Thus, the first two are represented as vectors, while the latter as a matrix. This class incorporates an Evaluate class, implementing a genotype-phenotype mapping. This mapping determines how each pair <feature,attribute> is encoded into representation of individuals.

\begin{figure}[htb]
    \centering
    \includegraphics[width=.48\textwidth]{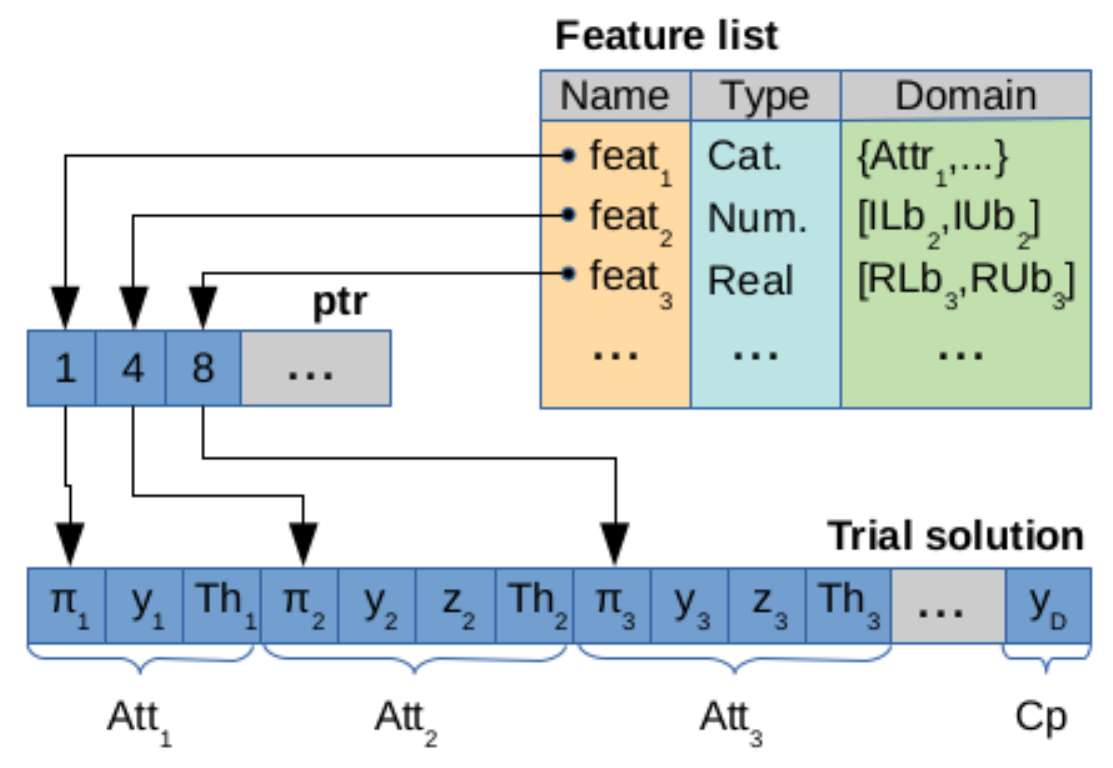}
    \caption{Genotype-phenotype mapping in the Evaluate class.}
    \label{fig:mapping}
\end{figure}

Indeed, a new nature-inspired algorithm serving as an extension of the existing framework needs to consider the following specifications: Each individual in the algorithm is encoded on the basis of the parsed feature list as a real-valued vector:
\begin{equation}
    \mathbf{x}_i=\{x_{i,1},\ldots,x_{i,D},x_{i,D+1}\},
\end{equation}
where $D$ is the problem size determined dynamically. Let us assume that the encoding depends on the feature type (Fig.~\ref{fig:mapping}). 
For instance, the features of a categorical type are encoded in the representation of a solution, using three elements that are decoded as:
\small
\begin{equation*}
    \textit{Attr}^{(\mathit{Cat})}_{\pi_j}=\left \{ \begin{array}{ll}
    x^{(\mathit{Cat})}_{i,\mathit{ptr}_{\pi_j}}\mapsto \pi_j, & \text{permutation of feature}, \\
    x^{(\mathit{Cat})}_{i,\mathit{ptr}_{\pi_j+1}}\mapsto y_{\pi_j}, & \text{categorical attribute}, \\
    x^{(\mathit{Cat})}_{i,\mathit{ptr}_{\pi_j+2}}\mapsto \mathit{Th}_{\pi_j}, & \text{threshold value}, \\
    \end{array}\right .
\end{equation*}
\normalsize
for $i=1,\ldots,\mathit{Np}$ and $j=1,\ldots,m$, where categorical value is calculated as:
\begin{equation}
    y_{\pi_j}=\lfloor x^{(\mathit{Cat})}_{i,\mathit{ptr}_{\pi_j+1}} \cdot |\mathcal{D}|\rfloor+1,
\end{equation}
and $|\mathcal{D}|$ is the size of the domain. The numeric features are encoded using even four elements decoded as:
\small
\begin{equation}
    \textit{Attr}^{(\mathit{Num})}_{\pi_j}=\left \{ \begin{array}{ll}
    x^{(\mathit{Num})}_{i,\mathit{ptr}_{\pi_j}}\mapsto \pi_j, & \text{permutation of feature}, \\
    x^{(\mathit{Num})}_{i,\mathit{ptr}_{\pi_j+1}}\mapsto y_{\pi_j}, & \text{integer lower bound}, \\
    x^{(\mathit{Num})}_{i,\mathit{ptr}_{\pi_j+2}}\mapsto z_{\pi_j}, & \text{integer upper bound}, \\
    x^{(\mathit{Num})}_{i,\mathit{ptr}_{\pi_j+3}}\mapsto \mathit{Th}_{\pi_j}, & \text{threshold value}, \\
    \end{array}\right .
\end{equation}
\normalsize
where the two middle elements encode an integer interval of feasible values $[y_{\pi_j},z_{\pi_j}]$ expressed as:
\small
\begin{equation*}
    y_{\pi_j}=\left \{ \begin{array}{ll} 
    \lfloor \left ( \textit{Ub}-\textit{Lb} \right ) x^{(\mathit{Num})}_{i,\mathit{ptr}_{\pi_j+1}} \rfloor, & \text{if}~x^{(\mathit{Num})}_{i,\mathit{ptr}_{\pi_j+1}}<x^{(\mathit{Num})}_{i,\mathit{ptr}_{\pi_j+2}}, \\
    \lfloor \left ( \textit{Ub}-\textit{Lb} \right ) x^{(\mathit{Num})}_{i,\mathit{ptr}_{\pi_j+2}} \rfloor, & \text{otherwise},
    \end{array} \right .
\end{equation*}
\normalsize
and
\small
\begin{equation*}
    z_{\pi_j}=\left \{ \begin{array}{ll} 
    \lfloor \left ( \textit{Ub}-\textit{Lb} \right ) x^{(\mathit{Num})}_{i,\mathit{ptr}_{\pi_j+2}} \rfloor, & \text{if}~x^{(\mathit{Num})}_{i,\mathit{ptr}_{\pi_j+1}}<x^{(\mathit{Num})}_{i,\mathit{ptr}_{\pi_j+2}}, \\
    \lfloor \left ( \textit{Ub}-\textit{Lb} \right ) x^{(\mathit{Num})}_{i,\mathit{ptr}_{\pi_j+1}} \rfloor, & \text{otherwise}.
    \end{array} \right .
\end{equation*}
\normalsize
On the other hand, the real-valued features are encoded similarly as numeric, except values $y_{\pi_j}$ and $z_{\pi_j}]$ are defined as integer in the first, and as float in the latter case. Additionally, the truncate operation is not necessary in the corresponding mapping calculations. 

Permutation $\Pi=(\pi_1,\ldots,\pi_m)$ is needed to modify ordering of the <feature,attribute> pairs within the association rules. Technically, all first elements in the corresponding feature representations are sorted in descendent order, 
\begin{equation*}
     x^{(\mathit{.})}_{i,\mathit{ptr}_{\pi_1}}\leq x^{(\mathit{.})}_{i,\mathit{ptr}_{\pi_2}}\leq\ldots\leq x^{(\mathit{.})}_{i,\mathit{ptr}_{\pi_m}},
\end{equation*}
while their ordinal values determine their position in the permutation. The threshold value denotes the presence or absence of the $\pi_j$-th feature in the observed association rule according to the following equation:
\begin{equation*}
    \mathit{Th}^{(.)}_{\pi_j}=\left \{ \begin{array}{lc}
    \mathit{enabled}, & \text{if}~\textit{rand}(0,1)< \begin{cases} x^{(\mathit{Cat})}_{i,\mathit{ptr}_{\pi_j+2}} \\ x^{(\mathit{Num|Real})}_{i,\mathit{ptr}_{\pi_j+3}}, \end{cases} \\
    \mathit{disabled}, & \text{otherwise}, \\
    \end{array}\right .
\end{equation*}
where $\textit{rand}(0,1)$ draws a value from uniform distribution in interval $[0,1]$.

Interestingly, the last element in the representation of the solution presents the so-called cutting point that determines the position of the implication sign in the association rule. This point is expressed as:
\begin{equation}
    \mathit{Cp}_i=\lfloor x_{i,D+1}\cdot (D-1)\rfloor +1,
\end{equation}
where $\mathit{Cp}_i\in[1,D-1]$. In summary, the length of the representation is $D+1$, and $D$ is calculated as:
\begin{equation}
    D=\sum^{m}_{j=1}{L\left(\mathit{Attr}^{(.)}_j\right)},
\end{equation}
where $L(.)$ designates the length of a definite feature.

After performing genotype-phenotype mapping, association rule $X\Rightarrow Y$ is decoded from a representation of an individual, where the quality of the association rule is evaluated using a fitness function. The fitness function is calculated according to the following equation:
\small
\begin{equation}
    f(\mathbf{x}^{(t)}_i)=\frac{\alpha\cdot\mathit{supp}(X\Rightarrow Y)+\beta\cdot\mathit{conf}(X\Rightarrow Y)+\gamma\cdot\mathit{incl}(X\Rightarrow Y)}{\alpha+\beta+\gamma},
    \label{eq:fit} 
\end{equation}
\normalsize
where $\alpha$, $\beta$, and $\gamma$ denote weights, $\mathit{supp}(X\Rightarrow Y)$ and $\mathit{conf}(X\Rightarrow Y)$ represent the support and confidence of the observed association rule, respectively, and $\mathit{incl}(X\Rightarrow Y)$ is defined as follows: 
\begin{equation}
    \mathit{incl}(X\Rightarrow Y)=\frac{\mathit{ante}(X\Rightarrow Y)+\mathit{cons}(X\Rightarrow Y)}{m}.
    \label{eq:incl}
\end{equation}
In Eq.~(\ref{eq:incl}), $\mathit{ante}(X\Rightarrow Y)$ represents a set of objects belonging to antecedent and $\mathit{cons}(X\Rightarrow Y)$ a set of objects belonging to consequent. Mathematically, these functions are expressed as: 
\begin{center}
$\mathit{ante}(X\Rightarrow Y) = \{o_{\pi_j}|\pi_j<\mathit{Cp}^{(t)}_i\wedge \mathit{Th}(\mathit{Attr}^{(t)}_{\pi_j}) = \mathit{enabled}\}$, \\
$\mathit{cons}(X\Rightarrow Y) = \{o_{\pi_j}|\pi_j\geq\mathit{Cp}^{(t)}_i\wedge \mathit{Th}(\mathit{Attr}^{(t)}_{\pi_j}) = \mathit{enabled}\}$.\\
\end{center}

The results of the optimization algorithm are stored into an archive of association rules. As opposed to the traditional algorithms for ARM, this archive is not generated uniformly, where each rule with support and confidence better than $S_{\mathit{min}}$ and $C_{\mathit{min}}$ is stored unconditionally; there only the rules that outperform the best fitness are stored into the archive. In this way, the size of the archive is reduced strongly.

In summary, a developer needs to be focused on the way the algorithm modifies the existing solution by including a new nature-inspired algorithm into a framework, while the evaluation component with complex genotype-phenotype mapping is already part of the framework.  

\subsubsection{VisualGuide class}
Visualization of mined association rules using the contemporary visual methods, like the Sankey diagram (also flow diagram) or Metro maps, are very complex, due primarily to the huge amount of rules. Therefore, a need for visual explaining the knowledge hidden in an archive has emerged. In line with this, algorithms for analyzing the archive have been developed that generate input data for visualization methods originally supported by external tools, like R, Matlab etc. 

The basis for analyzing the archive of association rules in uARMSolver presents a VisualGuide base class, illustrated in the listing below, that implements data structures and functions for handling this. 
\begin{lstlisting}
(*@\textbf{public}@*):
    (*@\text{VisualGuide}@*)((*@\text{Archive}@*) (*@\textit{arch}@*));
    (*@\text{$\sim$VisualGuide}@*)((*@\textbf{void}@*));
    (*@\textbf{virtual void}@*) (*@\text{Draw}@*)((*@\textbf{int}@*) (*@\textit{n}@*), (*@\textbf{int}@*) (*@\textit{m}@*));
\end{lstlisting}
Actually, this class defines a virtual function Draw(.) with parameters controlling the analysis. For instance, creating the Sankey diagram depends on the parameter map size that defines the maximum number of association rules observed according to a similarity measure. 

\section{Examples of use}~\label{sec_4}
This section is devoted to explaining how to install the framework and customize their tasks for best suiting the needs of specific users.

\subsection{Installation details}
Software in C++ programming code is available in the GitHub repository named \textit{uARMSolver} and available via permanent web link~\footnote{https://github.com/firefly-cpp/uARMSolver}. The framework is supported for both Windows and Linux operating systems. To install it on Linux directly from the source code, the following sequence of commands needs to be run in bash shell:

\begin{lstlisting}
(*@\text{\$ git clone https://github.com/firefly-cpp/uARMSolver}@*)
(*@\text{\$ cd uARMSolver}@*)
(*@\text{\$ make}@*)
\end{lstlisting}

The first command clones the GitHub repository with source code into a local directory \textit{uARMSolver}. The next command changes the current position to the local directory. The last command launches compiling and linking of the source and object code, respectively.

The framework can be launched using the following command:

\begin{lstlisting}
(*@\text{\$ uARMSolver -v}@*)
\end{lstlisting}
which displays the uARMSolver version, together with all the potential options of the command. Normally, the framework is executed using the following command:

\begin{lstlisting}
(*@\text{\$ uARMSolver -sarm.set}@*)
\end{lstlisting}
where the configuration file is also specified.

\subsection{Configuration example}
The configuration file (in our case ''arm.set'') allows customization of the framework environment. As said in Section~\ref{sec_3}, this file consists of three parts. Examples of the descriptions of all these parts are illustrated in the listings that follow.

The first configuration part defines the problem to be solved.
\begin{lstlisting}
(*@\text{\% Problem definition}@*)
(*@\text{Problem}@*)
{        
(*@\text{Tdbase\_name = datasets/Abalone.csv}@*)
(*@\text{Rule\_name = rules/Abalone.txt}@*)
(*@\text{Out\_name = out/Abalone.txt}@*)
(*@\text{Period = 1}@*)
}
\end{lstlisting}
Thus, names of input files (i.e., transaction database and archive of association rules) and output file (i.e., mined association rules) are specified. Parameter 'Period' indicates that only one transaction file ''Abalone.csv'' must be processed in this run of the framework.

The algorithm selection part is divided into two sections: The former is devoted for selecting the specific algorithm, while the latter for defining the proper values of parameters.

\begin{lstlisting}
(*@\text{\% Algoritem selection = {NONE, DE, PSO, ...}}@*)
(*@\text{Algorithm = DE}@*)
(*@\text{\% DE parameters}@*)
(*@\text{DE\_PARAM}@*)
{
(*@\text{DE\_NP = 100}@*)
(*@\text{DE\_FES = 1000}@*) 
(*@\text{DE\_RUNS = 1}@*)
(*@\text{DE\_F = 0.5}@*)
(*@\text{DE\_CR = 0.9}@*)
(*@\text{DE\_STRATEGY = 6}@*)
}
\end{lstlisting}
In our case, the DE algorithm was selected together with the corresponding parameters enclosed within 'DE\_PARAM' block.

The visualization method part is organized similar to the last one. However, the desired visualization method is selected in the first section, while parameters for controlling the algorithm for preparing data to be visualized are specified in the second.

\begin{lstlisting}
(*@\text{\% Visualization = {NONE. FLOW, METRO, ...}}@*)
(*@\text{Visualisation = FLOW}@*)
(*@\text{\% FLOW parameters}@*)
(*@\text{FLOW\_PARAM }@*)
{
(*@\text{FLOW\_M = 10}@*)
}
\end{lstlisting}
As can be seen from the listing, the Sankey diagram is selected that is controlled using one parameter 'FLOW\_M', indicating that the ten most similar rules are considered for visualization.

\section{Experiments}~\label{sec_5}
The purpose of the experimental section is to show how a work with the framework uARMSolver flows. In line with this, the Abalone transaction database is taken into consideration that is part of the UCI ML repository~\cite{uci2020repository}. The framework employed a parameter setting as proposed in the last section during the tests.

The Abalone dataset is dedicated for predicting the age of abalone from physical parameters. This prediction is normally performed in practice by using the shell through the cone, straining it and counting the number of rings under a microscope. This detaset incorporates 9 features, and contains 4,177 transactions. Its main advantage is that attributes are of all three types, i.e., categorical, numerical, and real-valued. Thus, the ability of the framework can be exposed the most.

The feature list obtained by parsing the Abalone dataset is illustrated in Table~\ref{tab:1},
\begin{table}[htb]
    \centering
    \caption{Feature list built by parsing the Abalone ML dataset.}
    \label{tab:1}
    \begin{tabular}{|C{1cm}|L{3cm}|L{4cm}|R{3cm}|}
    \hline
    Nr. & Feature & Type & Domain \\ \hline
    1 & Sex & CATEGORICAL & \{M,F,I\} \\
    2  & Length & REAL-VALUED & [0.075,0.815]  \\
    3  & Diameter & REAL-VALUED & [0.055,0.65]  \\
    4 & Height & REAL-VALUED & [0,1.13]  \\
    5 & Whole weight & REAL-VALUED & [0.002,2.8255]  \\
    6  & Shucked weight & REAL-VALUED & [0.001,1.488]  \\
    7  & Viscera weight & REAL-VALUED & [0.0005,0.76]  \\
    8  & Shell weight & REAL-VALUED & [0.0015,1.005]  \\
    9  & Rings & NUMERICAL & [1,29]  \\ \hline
    \end{tabular}
\end{table}
from which it can be seen that the framework is really capable of distinguishing type and determining the domain values of a particular feature automatically.

A structure of a parsed transaction database is depicted in Table~\ref{tab:2},
\begin{table}[htb]
    \centering
    \caption{Transaction database built by parsing the Abalone ML dataset.}
    \label{tab:2}
    \begin{tabular}{|C{1cm}|L{10cm}|}
    \hline
    Nr. & Transaction \\ \hline
    1 & (M,0.455,0.365,0.095,0.514,0.2245,0.101,0.15,15) \\
    2 & (M,0.35,0.265,0.09,0.2255,0.0995,0.0485,0.07,7) \\
    3 & (F,0.53,0.42,0.135,0.677,0.2565,0.1415,0.21,9) \\
    4 & (M,0.44,0.365,0.125,0.516,0.2155,0.114,0.155,10) \\
    5 & (I,0.33,0.255,0.08,0.205,0.0895,0.0395,0.055,7) \\
    6 & (I,0.425,0.3,0.095,0.3515,0.141,0.0775,0.12,8) \\
    7 & (F,0.53,0.415,0.15,0.7775,0.237,0.1415,0.33,20) \\
    8 & (F,0.545,0.425,0.125,0.768,0.294,0.1495,0.26,16) \\ \hline
    \end{tabular}
\end{table}
while a part of an archive of association rules (precisely 8) is presented in Table~\ref{tab:3} that includes antecedent and consequent parts of each association rule, to which also some measures, such as support, confidence, and fitness, are added. Due to the influence of the measure inclusion~\cite{Fister2020Inclusion}, the majority of attributes emerge in those rules having the higher fitness values. Let us notice that the total number of mined association rules in the archive was 337.

\begin{table*}[htb]
    \centering
    \caption{A part of an archive of association rules obtained by DESolver mining the Abalone ML dataset.}
    \label{tab:3}
    \begin{tabular}{|L{5.5cm}|L{5.5cm}|R{1cm}|R{1cm}|R{1cm}|}
    \hline
\multicolumn{1}{c|}{\textbf{Antecedent}}	&	\multicolumn{1}{c|}{\textbf{Consequent}}	&	\multicolumn{1}{c|}{\textbf{Supp.}}	&	\multicolumn{1}{c|}{\textbf{Conf.}}	&	\multicolumn{1}{c|}{\textbf{Fit.}}	\\ \hline
\makecell[l]{$0.5092\leq \text{Diameter} \leq 0.6500~\wedge$}  & \multirow{6}{*}{$0.9394\leq \text{Shucked weight}\leq 1.2169$} & \multirow{6}{*}{0.9998} & \multirow{6}{*}{1.0000} & \multirow{6}{*}{0.9629} \\ 
\makecell[l]{$0.3749\leq \text{Length}\leq 0.5928~\wedge$} & & & & \\
\makecell[l]{$16\leq \text{Rings}\leq 19~\wedge$} & & & & \\
\makecell[l]{$0.0015\leq \text{Shell weight}\leq 0.2999~\wedge$} & & & & \\
\makecell[l]{$0.4617\leq \text{Viscera weight}\leq 0.6136~\wedge$} & & & & \\
\makecell[l]{$0.0020\leq \text{Whole weight}\leq 1.8979$} & & & & \\ \hline
\makecell[l]{$0.3266\leq \text{Lenght}\leq 0.4572~\wedge$} & \makecell[l]{$0.3010\leq \text{Diameter} \leq 0.5219~\wedge$} & \multirow{4}{*}{1.0000} & \multirow{4}{*}{1.0000} & \multirow{4}{*}{0.9259} \\
$1\leq\text{Rings}\leq 13~\wedge$ & $0.8817\leq\text{Height}\leq 1.1300~\wedge$ & & & \\
$0.4616\leq\text{Shell weight}\leq 0.8758~\wedge$ & $0.3414\leq\text{Viscera weight}\leq 0.5737$ & & & \\
$0.5831\leq\text{Shucked weight}\leq 0.9894$ & & & &  \\ \hline 
\makecell[l]{$0.3836\leq\text{Height}\leq 0.9905~\wedge$} & $1\leq \text{Rings}\leq 29~\wedge$ & \multirow{4}{*}{0.9998} & \multirow{4}{*}{1.0000} & \multirow{4}{*}{0.9258} \\
$0.1040\leq\text{Length}\leq 0.5460~\wedge$ & $0.1034\leq\text{Shell weight}\leq 0.6760~\wedge$ & & & \\
$0.0005\leq\text{Viscera weight}\leq 0.4900~\wedge$ & $1.2177\leq\text{Shucked weight}\leq 1.4880$ & & & \\
$0.6163\leq\text{Whole weight}\leq 2.8255$ & & & &  \\ \hline
\multirow{5}{*}{\makecell[l]{$1.3163\leq\text{Shucked weight}\leq 1.3576$}} & \makecell[l]{$0.0550\leq\text{Diameter}\leq 0.1191~\wedge$}  & \multirow{5}{*}{1.0000} & \multirow{5}{*}{1.0000} & \multirow{5}{*}{0.8889} \\ 
 & $0.5946\leq\text{Height}\leq 0.7388~\wedge$ & & & \\
 & $4\leq \text{Rings}\leq 29~\wedge$ & & & \\
 & $0.2858\leq\text{Shell weight}\leq 0.7567~\wedge$ & & & \\
 & $0.0005\leq\text{Viscera weight}\leq 0.0819$ & & & \\
\hline
\end{tabular}
\end{table*}

\section{Conclusion}~\label{sec_6}
The ARM is one of the more interesting methods of ML that has been attracting many data scientists recently. A lot of traditional and nature-inspired algorithms have emerged for solving this problem. Typically, using these algorithms in practice is far from easy, because users are confronted with almost three problems: preprocessing the input data, association rule mining, and visualization of the mined rules. 

In order to help users, the paper proposes a framework for ARM named uARMSolver. The framework design is modular. Therefore, adding the new parser, the nature-inspired algorithm or visualization method is straightforward. It enables preprocessing transaction databases in different formats. Moreover, the parser is capable of determining types of attributes from their context automatically. Furthermore, the framework allows working with more transaction databases simultaneously. Nature inspired algorithms are applied for solving the ARM problem. These algorithms normally generate a lot of association rules, from which it is hard to discover the knowledge hidden in data. In order to overcome this problem, the framework is designed in the sense of EAI that tries to explain the knowledge in an archive of association rules using visualization. In line with this, the VisualGuide class was developed capable of analyzing the archive and incorporating these data to external tools for visualization (e.g., R, Matlab etc).

The uARMSolver was employed on Abalone UCI ML repository dataset. The results of DE for ARM showed that working with the framework is easy due to using the general setup file, while the obtained results can be different, depending on the used nature-inspired algorithm. This proves the fact that the results of the nature-inspired algorithm depend on the problem to be solved. Using more algorithms gives the framework an additional power.

There are several directions for improving the framework in the future. The most important direction is to connect the framework with different external visual tools. On the other hand, there are a lot of nature-inspired algorithms for ARM that deserve to be included into the framework.

\end{document}